\documentclass[aip,amsmath,amssymb,reprint]{revtex4-1}
\usepackage{graphicx}
\begin{document}
\title{Quantum Simulation of Phylogenetic Trees }
\author{Demosthenes Ellinas$^{\sharp}$ and Peter D. Jarvis$^{\diamondsuit}$}\mbox{}\\[.1cm]
\affiliation{
$^{\sharp}$ Technical University of Crete Department of Sciences \\
Math. Phys. and Quantum Information $M{\Phi}Q$ Research Unit
Chania Crete Greece \mbox{}\\[.1cm]
$^{\diamondsuit}$School of Mathematics and Physics, University of Tasmania, Australia\\
$^{\sharp}$ \texttt{\small ellinas@science.tuc.gr}
$^{\diamondsuit}$\texttt{\small Peter.Jarvis@utas.edu.au}}
%
\begin{minipage}[t]{6.0in}
\begin{abstract}
{\small Quantum simulations constructing probability tensors of biological multi-taxa in phylogenetic trees are proposed, in terms of positive trace preserving maps, describing evolving systems of quantum walks with multiple walkers. Basic phylogenetic models applying on trees of various topologies are simulated following appropriate decoherent quantum circuits. Quantum simulations of statistical inference for aligned sequences of biological characters are provided in terms of a quantum pruning map operating on likelihood operator observables, utilizing state-observable duality and measurement theory.}
\end{abstract}
\end{minipage}
\maketitle
%
%
\newtheorem{theorem}{Theorem}
\newtheorem{acknowledgement}[theorem]{Acknowledgement}
\newtheorem{algorithm}[theorem]{Algorithm}
\newtheorem{axiom}[theorem]{Axiom}
\newtheorem{claim}[theorem]{Claim}
\newtheorem{conclusion}[theorem]{Conclusion}
\newtheorem{condition}[theorem]{Condition}
\newtheorem{conjecture}[theorem]{Conjecture}
\newtheorem{corollary}[theorem]{Corollary}
\newtheorem{criterion}[theorem]{Criterion}
\newtheorem{definition}[theorem]{Definition}
\newtheorem{example}[theorem]{Example}
\newtheorem{exercise}[theorem]{Exercise}
\newtheorem{lemma}[theorem]{Lemma}
\newtheorem{notation}[theorem]{Notation}
\newtheorem{problem}[theorem]{Problem}
\newtheorem{proposition}[theorem]{Proposition}
\newtheorem{remark}[theorem]{Remark}
\newtheorem{solution}[theorem]{Solution}
\newtheorem{summary}[theorem]{Summary}
%
%
\textit{Introduction}: In the last two decades quantum mechanics has found
itself in a situation that could be characterized as an epistemological
exodus. It has expanded its scope and applicability into new fields, such as
information theory, the theory of computation, and even
biology, and has addressed fundamental problems and procedures of these
fields, by means of its physical-mathematical conceptual and
computational apparatus \cite{nc,qmbiol}. What were previously accepted as
quantum paradoxes and oddities, like quantum entanglement, have turned out
to be the keys to constructing novel computational and communicational
algorithms, providing the means for launching a new quantum technology. In
this vein, this paper puts forward a novel application of the discipline of
quantum computation-information to the field of evolutionary phylogenetics
\cite{ss,felbook}. Phylogenetics' main task is to construct ancestral
relationships (phylogenies), inferred by analyzing statistical data,
collected for various (morphological or genotypic) kinds of characters or
traits, possessed by selected groups of biological organisms (taxa). This
amounts to construction of phylogenetic trees with appropriate branching
patterns and evolutionary lengths, that successfully reproduce statistical
trends of alignments of sequences of certain characters \cite{ss,felbook}.
Various evolutionary models that compete by adjusting their tree vertex
transition probabilities, to accomplish this computationally NP-hard task
\cite{NPhard}, are then assessed by some statistical estimation such as maximum likelihood measure\cite{felbook}.

In this work a quantum simulation of phylogenetic evolution and inference,
is introduced in terms of trace preserving maps operating on quantum density
matrices. Basic multi-parametric evolutionary models are simulated, and
an association between phylogenetic trees and quantum circuits is
established. Specifically, group-based models are associated to quantum
walks (QW), and the Felsenstein model is shown to be related to
post-measurement state maps. Finally a quantum simulation of the iterative
pruning process for estimating maximum likelihood of phylogenetic trees, is
established in terms of quantum measurements of likelihood operator valued
measures (observables).

\textit{Notation}: Let the character set be $\Sigma =\{0,1,...,N\!-\!1\}=\{0\}\cup
\Sigma ^{\ast }$. Here $0$ is considered to be the \textquotedblleft null\textquotedblright\mbox{} or no character
symbol. Introduce the Hilbert space of character states $H\approx
l_{2}(\Sigma )=span(\left\vert i\right\rangle ;i\in \Sigma )$, of dimension $
dimH=\left\vert \Sigma \right\vert ,$ and consider the space $Lin(H)$ of
linear operators acting on $H$. Examples are the complete set of projectors $%
\widehat{P}_{i}=\left\vert i\right\rangle \left\langle i\right\vert ,$ $i\in
\Sigma ,$ the shift operator $h\left\vert i\right\rangle =\left\vert
i+_{N}1\right\rangle ,$ with $+_{N}$ addition modulo $N,$ (so that $%
h^{N}=1), $ and the space of density matrices ${\mathcal D}(H)\subset Lin(H)$.
A classical (discrete) probability distribution is represented as a vector $%
(p_{0},p_{1},p_{2},...p_{N-1}),$ and the corresponding quantum stochastic
system is represented by a diagonal density matrix $\rho =\sum_{i\in \Sigma
^{\ast }}p_{i}\widehat{P}_{i}\in \mathcal{D}(H);$ for biological
applications we will always assume $p_{0}=0$ (so that in practice the sum
runs only over characters $i\in \Sigma ^{\ast })$. On bipartite systems, the
unitary control-not operator $U_{cn}\in Lin(H\otimes H)$ defined as $%
U_{cn}=\sum_{k\in \Sigma }\widehat{P}_{k}\otimes h^{k},$ acts as\cite%
{nc} $%
U_{cn}\left\vert i\right\rangle \otimes \left\vert j\right\rangle
=\left\vert i\right\rangle \otimes \left\vert j+_{N}i\right\rangle. $

\textit{Splitting, cladogenesis, speciation}: The splitting operation \cite{splitting}
 $%
\Delta $ for given $1$-taxon matrix $\rho =\sum_{i\in \Sigma^{\ast } }p_{i}\widehat{P%
}_{i},$ is implemented by the adjoint action of $U_{cn}$
\begin{equation}
\Delta \rho =U_{cn}(\rho \otimes \widehat{P}_{0})U_{cn}^{\dagger
}=\sum_{i,j\in \Sigma^{*} }p_{ij}\widehat{P}_{i}\otimes \widehat{P}_{j},
\label{2taxa}
\end{equation}%
where $p_{ij}=p_{i}\delta _{ij},$ so $\Delta \rho $ is identified with a
two-taxon density matrix. The control-not gate embedded in various positions in $s$-fold products of character spaces, e.g. ${\mathbf 1}%
^{\otimes k-1}\otimes U_{cn}\otimes {\mathbf 1}^{\otimes s-k-1}$, provides the way to construct $s$-taxon phylogenetic trees of 
various topologies\cite{splitting}.

\begin{figure}
\scalebox{.6}{\includegraphics{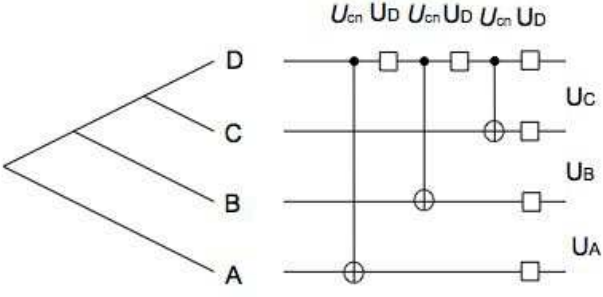}}
\caption{\label{fig:4taxa}A 4-taxa tree and its simulating quantum circuit.}
\end{figure}

\textit{Phyletic evolution, anagenesis}: For an $s$-taxon density matrix $\
\rho =\sum_{i_{1},...,i_{s}\in \Sigma^{*} }p_{i_{1}....i_{_{s}}}\widehat{P}%
_{i_{1}}\otimes ...\otimes \widehat{P}_{i_{s}},$ a suitable local unitary $%
U=\bigotimes_{i=1}^{s}U_{i}\in Lin(H)^{\otimes s},$ formalizes the phyletic
evolution\textit{\ }of taxa, when its action is composed with the $s$-fold
product of the local \textit{diagonalizing map} ${\mathcal E}_{d}^{\otimes s}$%
, where ${\mathcal E}_{d}(\cdot)=\sum_{k\in \Sigma }\widehat{P}_{k}(\cdot)\widehat{P}%
_{k},$ is the completely positive trace preserving (CPTP) map that projects
out the diagonal part of a matrix\cite{kraus}, that is a decoherent map. Thus we have $\rho
\rightarrow \widetilde{\rho }\equiv {\mathcal E}_{d}^{\otimes s}(U\rho
U^{\dagger })$, where
\begin{eqnarray}
\widetilde{\rho }&=&\!\!\!\sum_{i_{1},...,i_{s}\in \Sigma^{*} }\widetilde{p}%
_{i_{1}....i_{s}}\widehat{P}_{i_{1}}\otimes ...\otimes \widehat{P}_{i_{s}},
\label{s_taxa_rho} \\
\widetilde{p}_{i_{1}....i_{s}}&=&\!\!\!\sum_{j_{1},...,j_{s}\in \Sigma^{*}
}p_{j_{1}....j_{s}}(M_{1}\otimes ...\otimes
M_{s})_{i_{1}j_{1};...;}{}_{i_{s}j_{s}}. \qquad \mbox{} \label{pprime_2}
\end{eqnarray}
Abbreviating the adjoint action of an operator as $Ad\,S(\cdot)\equiv S(\cdot)S^{\dagger
},$ we say that the map ${\mathcal E}_{d}^{\otimes s}(Ad\,U)$ \ thus
induces a general doubly-stochastic transformation in the probability
tensor. The Hadamard or entry-wise product of matrices defined as $%
(A\circ B)_{ij}=A_{ij}B_{ij}$, has been used, to obtain\cite{maolkin} the
Markov matrices $M_{i}=U_{i}\circ U_{i}^{\ast }$, which will drive evolution
on edges of a model phylogenetic tree. Below, we make particular choices of $%
U$ \ to reflect different types of phylogenetic models.
Fig.~\ref{fig:4taxa} summarizes the preceding discussion by showing a four taxon tree and its simulating quantum circuits.

%
%
%
%

\textit{Phyletic evolution and quantum walks}: It has long been
appreciated that faithful modeling of trait evolution in phylogenetics is
problematic. As has been remarked, \textquotedblleft...Brownian motion is a poor model, and
so is Ornstein-Uhlenbeck\textquotedblright\mbox{}\cite{felmessage}. We here present a novel
proposal for the stochastic phyletic evolution of traits via quantum
simulation employing QWs (see \cite{qrw}), operating locally on density
matrices along edges of trees. This is set up as follows. Introduce in
additional to character Hilbert space $H$ \ (the \textquotedblleft walker\textquotedblright\mbox{} space), at each
node of phylogenetic tree an auxiliary \textquotedblleft coin\textquotedblright\mbox{} Hilbert space $H_{c}\approx l_{2}(C)=span(\left\vert +\right\rangle
,\left\vert -\right\rangle )$, and projectors $P_{\pm }\in
Lin(H_{c})$. Evolution now proceeds on joint \textquotedblleft
walker\textquotedblright\mbox{} and \textquotedblleft coin\textquotedblright\mbox{}
states $\rho _{c}\otimes \rho $ via a standard QW conditional unitary
operator $V=(P_{+}\otimes h+P_{-}\otimes h^{\dagger })U\otimes {\mathbf 1}$, acting from $H_{c}\otimes H$ to itself. One \textquotedblleft step\textquotedblright\mbox{} of such a QW is realized by the map
on the \textquotedblleft walker\textquotedblright\mbox{} density matrix,  viz.
$\rho \rightarrow
{\mathcal E}_{V^{k}}(\rho ):=Tr_{c}V^{k}(\rho _{c}\otimes \rho )V^{\dagger k},$
followed by diagonalization with ${\mathcal E}_{d}$. For $s$ taxa, $%
E_{V^{k}}\equiv ({\mathcal E}_{d}^{\otimes s}\circ {\mathcal E}%
_{V^{2}}^{\otimes s})$. For example for the two-taxon case, with $k=2$ and
coin initially in a pure state $\rho _{c}=\left\vert c\right\rangle
\left\langle c\right\vert $ with $|c\rangle=|+\rangle$ or $|-\rangle$, we obtain $%
E_{V^{2}}(\rho )=\sum_{mn}\widetilde{p}_{mn}\widehat{P}_{m}\otimes \widehat{P%
}_{n},$ with components $\widetilde{p}_{mn}=\sum_{ab}p_{m-a,n-b}q_{a}^{(c)}q_{b}^{(c)},$
where $q_{a}^{(c)}:=\sum_{\gamma }M_{\gamma ,a-\gamma }M_{\gamma -a,c}\geq
0, $ is a probability distribution (that is, $q_{a}^{(c)}>0,%
\sum_{s}q_{a}^{(c)}=1),$ determined by the coin tossing unitary $U$ via the
Hadamard product $M=U\circ U^{\ast }$. The tensor $\widetilde{p}$ so obtained, and its
multi-taxa generalizations, are objects of quantum simulations. Also the
diagonalizing map ${\mathcal E}_{d}$ can be cast in the form of a CPTP map, i.e. $%
{\mathcal E}_{d}(\rho )=\sum_{k\in \Sigma^{*} }\widehat{P}_{k}\rho \widehat{P}%
_{k}=\sum_{k\in \Sigma^{*} }q_{k}U_{k}\rho U_{k}^{\dagger }$ with each $q_{k}=%
{1}/{\left\vert \Sigma \right\vert },$ thanks to the non-uniqueness of
the operator sum representation, with unitaries $U_{k}$ related to
projectors by discrete Fourier transform, $U_{k}=\sum_{l}\omega ^{kl}\widehat{%
P}_{l}$ and $\omega =\exp (i2\pi /\left\vert \Sigma \right\vert )$. Below,
similar quantum prescriptions will be given to the structural maps of
standard evolutionary models.

\textit{Phylogenetic evolutionary models and quantum maps}: Next we exploit
the above considerations in specific cases of standard phylogenetic
models, namely the so-called group-based models (see references\cite{jk}): Jukes-Cantor (JC), Kimura
two-parameter (K2), Kimura three-parameter (K3), and the binary symmetric
model (B), as well as the Felsenstein model (F)\cite{jk}.
Firstly we give in each case a direct Kraus representation of the quantum
map $E_{\tau }\equiv {\mathcal E}_{d}\circ {\mathcal E}_{\tau }$. This is
followed by a QW formulation using, as above, an additional ancillary
\textquotedblleft coin\textquotedblright\mbox{} space. Let $X$, $Z$ denote the usual
single qubit not and phase gates (the Pauli matrices $\sigma _{x}$, $\sigma_z$ respectively) and $U_{kl}=$
$X^{k}\otimes X^{l}$, for $k,l = 0,1$. The following propositions are verified by
direct calculation for operators in $l_2 (\Sigma^{*})$ acting on 
$\rho =\sum_{m\in \Sigma^{*}
}p_{m}\widehat{P}_{m}$: \\

\noindent
\textit{Proposition K}: Let $\left\vert \Sigma^{*} \right\vert =4$ and $\tau \in \{K3,K2,JC\}$.  We have 
\begin{eqnarray}
E_{\tau }(\rho ) &=&\sum_{k,l}\lambda _{kl}^{(\tau )}U_{kl}(\rho
)U_{kl}^{\dagger }=\sum_{m\in \Sigma ^{\ast }}(M_{\tau }p)_{m}\widehat{P}%
_{m}, \nonumber \\
M_{\tau }(a,b,c)&=&\sum_{kl}\lambda _{kl}^{(\tau
)}U_{kl}\circ U_{kl}^{\ast }=\sum_{kl}\lambda _{kl}^{(\tau )}X^{k}\otimes
X^{l}.  \label{evolucptmm}
\end{eqnarray}%
The weights $\lambda_{k l}^\tau$ and corresponding model Markov matrices $M_\tau$ are defined as follows. For
generic parameters define the weights $\lambda _{kl}(a,b,c)$ as $\lambda _{00}=1\!-\!a\!-\!b\!-\!c,$ $\lambda _{10}=a,$ $\lambda
_{01}=b$, $\lambda _{11}=c$, and take the corresponding convex sum $M(a,b,c)$. Then $\lambda _{kl}^{(3K)}=\lambda _{kl}(a,b,c)$, $M_{3K}\equiv
M(a,b,c)$, $\lambda _{kl}^{(2K)}=\lambda _{kl}(a,b,b)$, $M_{2K}\equiv
M(a,b,b)$, and finally $\lambda _{kl}^{(JC)}=\lambda _{kl}(a,a,a)$, $M_{JC}\equiv
M(a,a,a)$.
\hfill $\Box$ \\

\noindent
\textit{Proposition K$\,'$:} \ The CPTP map $E_{\tau }$ has, in addition
to the operator sum representation above, also a QW like representation $%
E_{\tau }(\rho )=Tr_{c}V_{\tau }(\rho _{c}\otimes \rho )V_{\tau
}{}^{\dagger }$, in terms of a unitary dilation $V_{\tau
}=(\sum_{kl}P_{k}\otimes P_{l}\otimes U_{kl})U_{\tau }\otimes \mathbf{1}$
which acts on a composite coin-walker space $H_{c}\otimes H$, with
four-dimensional ancillary space. Here $V_{\tau }$ is a control-control-$%
U_{kl}$ operator. For a coin density matrix with spectral decomposition $\rho
_{c}=\sum_{k}\mu _{k}\left\vert c_{k}\right\rangle \left\langle
c_{k}\right\vert ,$ the coin-tossing unitary $U_{\tau }$ should
satisfy $\left\langle kl\right\vert U_{\tau }\circ U_{\tau }^{\ast
}\left\vert c\right\rangle =\lambda _{kl}^{(\tau )}$, with $\left\vert
c\right\rangle =\sum_{k}\mu _{k}\left\vert c_{k}\right\rangle $ a
stochastic vector. Also $U_{kl}=e^{i{\mathcal H}_{kl}}$ where ${\mathcal H}%
_{kl}=\frac 12{\pi }[-(k+l){\mathbf 1}\otimes {\mathbf 1}+kX\otimes {\mathbf 1}%
+l{\mathbf 1}\otimes X].$ 
\hfill $\Box$\\

\noindent
\textit{Proposition B:} Let $\left\vert \Sigma^{*} \right\vert =2$. The map ${\cal E}_d \circ E_{B}$, where $E_{B}(\rho
)=(1-a )\rho + a X\rho X^{\dagger }$, simulates the binary symmetric
model $M_{B}(a )$ $=(1-a )\mathbf{1+}a X$ $\ $acting
as $\rho =\sum_{m\in \Sigma^{*}}p_{m}\widehat{P}_{m}\rightarrow \sum_{m\in
\Sigma^{*} }(M_{B}\,p)_{m}\widehat{P}_{m}$. $\ \ \ $ 
\hfill $\Box$\\

\noindent
\textit{Proposition B$\,'$:} The \textquotedblleft control flip\textquotedblright\mbox{} map $E_{B}$ is
unitarized in composite coin-walker space with a two-dimensional ancillary
space as, $E_{B}(\rho )=Tr_{c}V_{B}(\rho _{c}\otimes \rho )V_{B}^{\dagger }$,
with the starting coin state $\rho_{c}=\left\vert 1\right\rangle \left\langle
1\right\vert $, and $V_{B} = \sqrt{{\scriptsize a }}\mathbf{1}\!\otimes\! \mathbf{1} +
\sqrt{{\scriptsize 1\!-\!a}} {\scriptsize Y}\!\otimes\! {\scriptsize X}$, and 
${\scriptsize Y}\equiv{\scriptsize Z}{\scriptsize X}$. 
\hfill $\Box$\\

%
\noindent
\textit{Remark}:
In the QW picture, the weight parameters $\lambda _{kl}^{(\tau )}$ determine non-uniquely, via the unistochastic\cite{unistochastic} matrix $U_{\tau
}\circ U_{\tau }^{\ast }$, the coin-tossing matrix $U_{c}$,
which in turn determines the $U$-quantization of the underlying classical walk\cite{qrwellsmyrn} with
evolution matrix $V_{cl}\equiv \sum_{kl}P_{k}\otimes P_{l}\otimes
U_{kl}$.

For the Felsenstein model (F)\cite{jk}, quantum
simulation requires the following ingredients. The model's stationary
distribution $(\pi _{1},\pi _{2},\pi _{3},\pi _{4})$, $\sum \pi _{i}=1$, is
to be used to introduce the observable ${\mathbf 1}_{\pi }:=4\sum_{i}\pi _{i}\widehat{P}%
_{i}$, with Kraus operators $F_{ij}=\sqrt{\pi _{j}}\left\vert i\right\rangle
\left\langle j\right\vert ,$ $i,j\in \Sigma =\{1,2,3,4\}$ obeying the resolution relation $%
\sum_{ij}F_{ij}^{\dagger }F_{ij}=\frac{1}{4}{\mathbf 1}_{\pi }$. Again let $\rho =\sum_{m\in \Sigma^{*} }p_{m}\widehat{P}_{m}$. By direct
calculation we obtain: \\

\noindent
\textit{Proposition F:} The quantum map implementing the Felsenstein model $\rho \rightarrow
E_{F}(\rho )=\sum_{m\in \Sigma^{*} }(M_{F}\,p)_{m}\widehat{P}_{m}$ is given by
\begin{equation}
E_{F}(\rho )=(1-a)\frac{1}{p_{\pi }}\sum_{i,j}F_{ij}\rho F_{ij}^{\dagger
}+a\rho \mathbf{,}
\end{equation}%
where $p_{\pi }=Tr(\sum_{i,j}F_{ij}^{\dagger }F_{ij}\rho )=Tr(\frac{1}{4}%
\mathbf{1}_{\pi }\rho )$ is a normalization constant, and the model's stochastic
matrix is obtained as $M_{F}=(1-a)\sum_{i,j}F_{ij}\circ F_{ij}+a\mathbf{1}%
$.

In the framework of quantum measurement theory, simulation of
the Felsenstein model is interpreted as follows. There are two observables: $%
{\mathbf 1}_{\pi }$ as above, and also ${\mathbf 1}_{\pi }^{\#}$
defined analogously in terms of the complementary probability distribution ($%
\pi_{1}^{\#},\pi_{2}^{\#},\pi_{3}^{\#},\pi
_{4}^{\#})$, with ${\mathbf 1=1}_{\pi }$ $+{\mathbf 1}_{\pi }^{\#}$ forming a non-orthogonal decomposition of unity. These
observables are measured by means of the so called instruments \cite{kraus},
which are the two families of Kraus generators: the $\{F_{ij}\}_{i,j=0}^{1}$
as above, and the analogous $\{F_{ij}^{\#}\}_{i,j=0}^{1}$ defined
in terms of $\pi^{\#}$ rather than $\pi $ (see e.g. \cite%
{kraus}). The measurement probabilities of the observables ${\mathbf 1}%
_{\pi }$ and ${\mathbf 1}_{\pi }^{\#}$ in the system are $p_{\pi }^{\#}=Tr({\mathbf 1}_{\pi }^{\#}\rho ),$ and $p_{\pi }=Tr(%
{\mathbf 1}_{\pi }\rho )$, and the action of quantum map $E_{F}$ on the density matrix $E_{F}(\rho )$
gives the post-measurement density matrix for a \textit{non-efficient \ }%
quantum measurement for observable ${\mathbf 1}_{\pi }$ of
\textit{finite strength }\cite{fuchs}. The complementary measurement of $%
\mathbf{1}_{\pi }^{\#}$ is not used. In the uniform
limit $\pi _{j}=\frac{1}{4}$ then ${\mathbf 1}_{\pi }={\mathbf 1,}$ ${\mathbf 1}%
_{\pi }^{\#}={\mathbf 0}$ and $p_{\pi }=1,$ and the model reduces
to the JC model.

\textit{Quantum estimation of likelihood}:
Our general framework also
encompasses the quantum estimation of model-based tree likelihoods 
 (F)\cite{jk}, whose numerical calculation and optimization provides a major tool for
phylogenetic inference (for computational
heuristics see e.g. \cite{RAxML}). Likelihood evaluation has been demonstrated
to be a computationally NP-hard problem \cite{NPhard}, and it is therefore
desirable to put forward a quantum simulation equivalent.
In the usual formulation (F)\cite{jk}, likelihood vectors are
initialized at the pendant nodes (leaves) of a tree, and are then computed recursively back to the root node,
the final result being a scalar quantity, the tree likelihood. The key operation is that of pruning,
that is, of arriving at the likelihood for a parent node, say $A$, by
combining a pair of daughter likelihoods, say $B$, $C$, from nodes which
root two sub-trees. Explicitly, likelihoods for daughter nodes $B$, $C$ are
combined to give the parent likelihood $L_{i}^{A}=(\sum_{j\in \Sigma ^{\ast
}}M_{ij}^{B}(t_{B})L_{j}^{B})(\sum_{k\in \Sigma ^{\ast
}}M_{ik}^{C}(t_{C})L_{k}^{C})$, where $M^{B,C}(t_{B,C})$ are stochastic
matrices depending on branch lengths $t_{B,C}$ specified by the evolutionary
model employed.  Next, an alignment of $s$ taxa over $\Lambda$ sites is
considered. If the characters at site $l$ of the alignment are $%
i_{1}^{(l)}i_{2}^{(l)}...i_{s}^{(l)}$, then likelihoods for the tips of the tree
(leaf nodes) are initialized to $L_{k}^{(l)}=\delta({k,i_{k}^{(l)}})$.
The pruning map is applied recursively at all cherries, and then higher up the
tree, to arrive at the total tree likelihood $L_{tr}^{(l)}=(L_{tr\text{ }%
k}^{(l)})_{k=1}^{s}$, which is finally averaged over the assumed stationary
distribution $(\pi _{i})$ of the model to obtain site $l$'s likelihood $%
L^{(l)}=\sum_{k}\pi _{i}L_{tr\text{ }k}^{(l)}$. For the entire alignment,
the tree (log) likelihood is then $L(T;w^{\ast })$ $=\max_{w}\log \Pi
_{l=1}^{\Lambda}$ $L^{(l)}$, where $T$ denotes the tree topology and $w^{\ast }$ \
the optimal model (weight) parameters.


\begin{figure}
\scalebox{.55}{\includegraphics{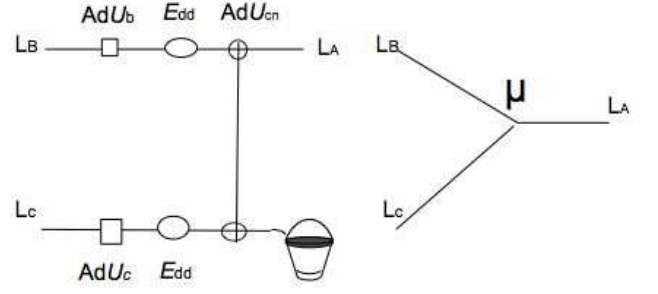}}
\caption{\label{fig:pruning} Circuit for pruning map of likelihood operators.}
\end{figure}

In the quantum simulation introduced here, likelihoods are regarded as
quantum observables, that is operators in $Lin(H)$, dual to density operators
under the trace inner product (see above). The likelihood operator at node $A$ has
components $\widehat{L}_{i}^{A}\equiv L^{A}(t|i)={\mathbb P}(i|t),$ where ${\mathbb P}(i|t)$ is
the conditional probability of character $i\in \Sigma ^{\ast },$ for
parameters $t=(T,w).$ Here $A=1,2,...,s$ are leaf nodes and $A=s+1,...,2s-2,$
internal (ancestral) nodes. Consider parent and daughter nodes $A$, $B$ and $C$,
with respective likelihood operators $\widehat{L}^{A}$, $\widehat{L}^{B}$ and 
$\widehat{L}^{C}$. Operators for daughter nodes $B$, $C$ are
combined using the analog of pruning, the quantum pruning map $\mu
:Lin(H)\otimes Lin(H)\rightarrow Lin(H)$ that provides the parent 
operator $\widehat{L}^{A}=\mu (\widehat{L}^{B}\otimes \widehat{L}^{C}),$
where $\mu =Tr_{B}\circ Ad\,U_{cn}^{\dagger }\circ {\mathcal E}_{dd}\circ
Ad(U_{B}\otimes U_{C})$. The map $\mu $ uses stochastic matrices $%
M^{x}(t_{x})=U_{x}\circ U_{x}^{\ast },$ depending on branch lengths $t_{x}$
for $x=A,B,$ as given by the model employed, and the collective
\textquotedblleft diagonalizing map\textquotedblright\mbox{} ${\mathcal E}_{dd}(\cdot)=\sum_{k}\widehat{P}_{k}\otimes
\widehat{P}_{k}(\cdot)\widehat{P}_{k}\otimes \widehat{P}_{k}. $ Fig.~\ref{fig:pruning}
presents a quantum circuit realizing map $\mu$. By using its
embedding $\mu_{r,r+1}=id^{\otimes r-1}\otimes \mu \otimes id^{\otimes
s -r}$ for various values of $r$ according to the topology of the
binary tree, the pruning map $\mu $ is applied recursively to all cherries, and
then higher up the tree. In this way we arrive at the tree likelihood operator
$\widehat{L}_{tr}^{(l)}$, which then is contracted with model's stationary
density matrix $\rho ^{\pi }=\sum_{i}\pi _{i}\widehat{P}_{i},$ to yield as a
measurement result the site $l$ likelihood $L^{(l)}=Tr(\widehat{L}%
_{tr}^{(l)}\rho ^{\pi })\equiv \left\langle \widehat{L}_{tr}^{(l)},\rho
^{\pi }\right\rangle .$ For the entire alignment, the tree (log)
likelihood is (c.f. the identity $Tr(AB)Tr(CD)=Tr(A\otimes C)(B\otimes D)$ )%
\begin{equation*}
L=\max_{w}\log \prod_{l=1}^{\Lambda}\left\langle \widehat{L}%
_{tr}^{(l)},\rho ^{\pi }\right\rangle =\max_{w}\log Tr(\otimes
_{l=1}^{\Lambda}\widehat{L}_{tr}^{(l)})\rho_{\Lambda }),
\end{equation*}%
where $\rho_{\Lambda}\equiv (\rho^{\pi})^{\otimes \Lambda}$ is the product
of $\Lambda$ stationary density matrices. 

In fact this Heisenberg-like
picture of updating the observables (likelihoods), and finally contraction with
the stationary density matrix to derive site and eventually alignment
likelihoods, can be converted to a Schr\"{o}dinger-like picture, using the
observable-state duality, exemplified here by the trace cyclic property.
Firstly note that the pruning map can be expressed as $\mu (\widehat{L}{}%
^{B}\otimes \widehat{L}{}^{C})=\nu^{-1}{\mathcal E}_{B}(\widehat{L}{}^{C}),$ $%
\nu =Tr\widehat{L}{}^{B},$ where the positive stochastic map ${\mathcal E}_{B}$
decomposes as ${\mathcal E}_{B}\equiv {\mathcal E}_{Bpd}\circ Ad\,U_{C}$ \ with $%
{\mathcal E}_{Bpd}(\cdot)=\sum_{k\in \Sigma }q_{k}^{B}Ad\,\widehat{P}_{k}(\cdot),$ \ a
probabilistic diagonalizing map, with probabilities $q_{k}^{B}=\nu^{-1}%
\left\langle k\right\vert U_{B}\widehat{L}{}^{B}U_{B}^{\dagger }\left\vert
k\right\rangle .$ As the roles of $\widehat{L}{}^{B}$ and $\widehat{L}{}^{C}$
can be exchanged above with appropriate modification, (${\mathcal E}_{B}$
becomes ${\mathcal E}_{C}$ etc), we note that $\mu $ is proportional to a
stochastic map either way, and by duality it can be made to act on
density matrices instead of likelihood operators. This is also true for
embedded pruning maps $\mu _{r,r+1}$, i.e. they will also be proportional
to maps ${\mathcal E}_{B;r,r+1}$ for the appropriate current likelihood $\widehat{L}{}^{B}$
etc. Then the tree likelihood operator $\widehat{L}_{tr}^{(l)},$ obtained by
composing pruning maps, will eventually be described by pruning a final
cherry, say with nodes $B_{f}$ and $C_{f},$ ie. $\widehat{L}_{tr}^{(l)}=%
{\nu _{f}}^{-1}{\mathcal E}_{B_{f}}(\widehat{L}{}^{C_{f}}),$ $\nu _{f}=Tr%
\widehat{L}{}^{B_{f}}.$ Then the likelihood at site $l$ is obtained as $L^{(l)}={\nu _{f}}^{-1}
Tr({\mathcal E}_{B_{f}}(\widehat{L}{}^{C_{f}})\rho ^{\pi })={\nu _{f}}^{-1}
Tr(\widehat{L}{}^{C_{f}}{\mathcal E}_{B_{f}}^{\ast }(\rho ^{\pi })),$
where the dual map ${\mathcal E}_{B_{f}}^{\ast }$ of ${\mathcal E}_{B_{f}}$
acting on the density matrix is introduced. This situation is extended  similarly to the
likelihood of the entire alignment by assigning additional site indices $l$
to each likelihood operator, e.g. $\widehat{L}{}_{l}^{B_{f}}$ and $\widehat{L}{}%
_{l}^{C_{f}}$, as well as trace coefficients $\nu _{f}^{l}$ etc, to obtain $L(T;w^{\ast
})=\max_{w}\log \prod_{l=1}^{\Lambda}(\nu _{f}^{l})^{-1}Tr(\widehat{L}%
_{\Lambda }(\otimes _{l=1}^{\Lambda }{\mathcal E}_{B_{f;l}}^{\ast })\rho
_{\Lambda })$. Here $\widehat{L}_{\Lambda }\equiv \otimes _{l=1}^{\Lambda }(%
\widehat{L}_{l}^{C_{f}})$ is the product of $\Lambda $ different likelihood
operators, corresponding to final cherries of the respective trees, employed to
construct tree likelihoods. Note that $\otimes _{l=1}^{\Lambda }{\mathcal E}%
_{B_{f;l}}^{\ast }$ is a collective factorized map that can be expressed in
terms of a unitary dilation, and this would in principle be implemented by a Hamiltonian
quantum model.

In conclusion, this study lays the groundwork for simulating, by quantum
mechanical means, the probability tensors of multi-taxa systems, and for
estimating the maximal likelihood of a phylogenetic alignment. With the
tools developed here, prominent among problems for future investigations
would be for example a quantum computational simulation of Steel's conjecture
\cite{Sconjecture} and its resolution\cite{scProof}.

\textit{Acknowledgements:}
PDJ thanks the Technical University of Crete Department of Sciences, and 
Mathematical Physics and Quantum Information $M{\Phi}Q$ Research Unit for hospitality during a collaborative visit. Similar appreciation is expressed to 
the Australian-American Fulbright Foundation, and 
staff and colleagues at the Department of Statistics, University of California Berkeley, as well as the Department of Physics, University of Texas Austin, for visits as an Australian senior Fulbright scholar, during part of this work.
%

%


\end{document}